\pdfoutput=1
\documentclass[aps,prl,twocolumn]{revtex4} 
\usepackage{graphicx}
\usepackage[utf8]{inputenc}

\newcommand{\un}[1]{\ensuremath{\,\textrm{#1}}}
\newcommand{\vsd}{\ensuremath{V_{\textrm{sd}}}}
\newcommand{\vg}{\ensuremath{V_{\textrm{g}}}}
\newcommand{\didv}{\ensuremath{\textrm{d}I/\textrm{d}\vsd}}

\begin{document}

\title{Negative frequency tuning of a carbon nanotube nano-electromechanical 
resonator}

\author{P. L. Stiller}
\author{S. Kugler}
\author{D. R. Schmid}
\author{C. Strunk}
\author{A. K. H\"uttel}
\affiliation{Institute for Experimental and Applied
Physics, University of Regensburg, 93040 Regensburg, Germany}

\begin{abstract}
A suspended, doubly clamped single wall carbon nanotube is characterized as 
driven nano-electromechanical resonator at cryogenic temperatures. 
Electronically, the carbon nanotube displays small bandgap behaviour with 
Coulomb blockade oscillations in electron conduction and transparent contacts 
in hole conduction. We observe the driven mechanical resonance in dc-transport, 
including multiple higher harmonic responses. The data shows a distinct negative frequency tuning at finite applied gate 
voltage, enabling us to electrostatically decrease the resonance frequency to 
75\% of its maximum value. This is consistently explained via electrostatic 
softening of the mechanical mode.
\end{abstract}

\maketitle   

\section{Introduction}
As has been shown in many both room temperature 
\cite{nature-sazonova:284,nl-witkamp:2904} and low temperature measurements 
\cite{highq,strongcoupling,magdamping}, suspended carbon nanotubes display 
excellent properties as mechanical resonator systems. Low-temperature 
measurements have displayed mechanical quality factors up to $Q\sim 10^5$ 
\cite{highq,HighQuality} at frequencies in the megahertz to gigahertz 
range. Multiple higher harmonics have been observed with frequencies up to 
$39\un{GHz}$ \cite{doi:10.1021/nl203279v}. One additional feature of particular 
interest of carbon nanotube nano-electromechanical resonators is the high 
tunability of the resonance frequency: the particular combination of high 
Young's modulus \cite{prl-lu:1297,prl-yu:5552}, low diameter and low mass make 
it possible to tune a carbon nanotube ``beam resonator'' all the way from the 
case of a hanging chain to a rope with high tensile load 
\cite{book-cleland,nl-witkamp:2904} by applying electrostatic forces via gate 
voltages alone. As expected from the bulk beam model, the resonance frequency 
typically has a minimum around zero applied gate voltage, and increases at 
finite voltage. We report here on a particular resonator where we have observed 
a strong manifestation of the opposite behaviour: the resonance frequency can 
be tuned down to $\sim 75\%$ of its maximum value.

\section{Sample fabrication}
Our devices are fabricated using standard lithography techniques. Base material 
is a highly positive doped silicon wafer with a thermally grown $550\un{nm}$ 
silicon dioxide layer on top. 
\begin{figure*}[htb]%
\includegraphics*[width=\textwidth]{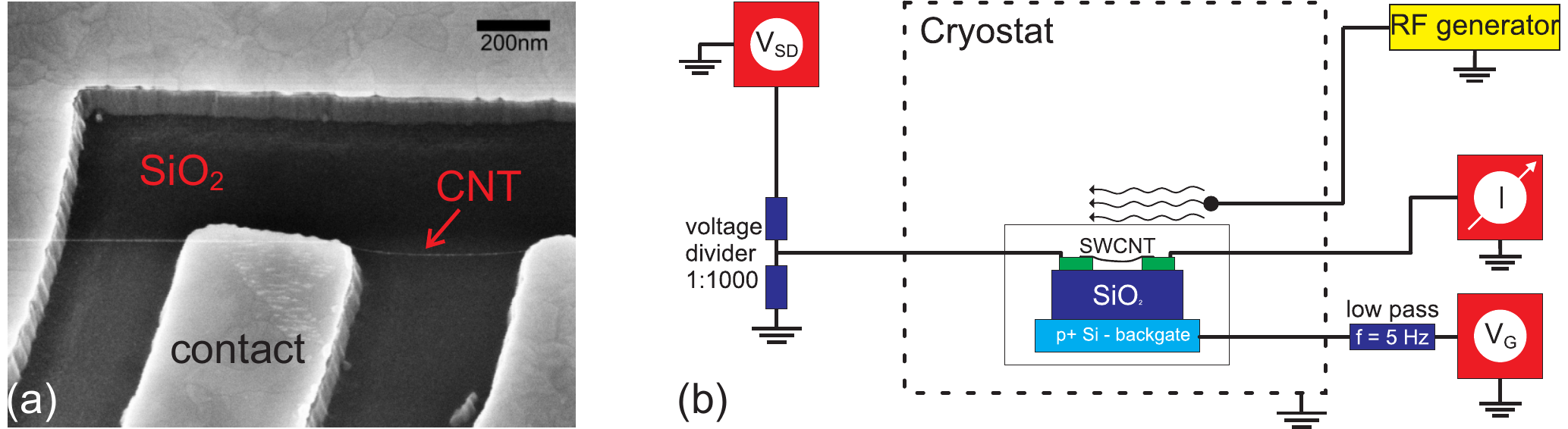}
\caption{(a) Detail SEM image of a typical suspended carbon nanotube sample: a 
suspended nanotube lies across metallized contact electrodes and the trenches 
etched between them. (b) Schematic of the low temperature electronic 
transport measurement setup; as cryostat, an Oxford Instruments helium-3 
evaporation system was used.} 
\label{probe}
\end{figure*}
Figure \ref{probe}(a) displays a typical detail SEM image of the chip geometry 
used in our measurements. We achieve suspended and contamination-free carbon 
nanotubes by chemical vapor deposition growth over predefined contacts 
separated by etched trenches \cite{nmat-cao:745,nnano-steele:363,magdamping}. 
As contact material, we use a 40\,nm thick, co-sputtered layer of a rhenium / 
molybdenum alloy \cite{magdamping,Schneider}. This material also serves as etch 
mask for the trench definition. Various geometries are used in different chip 
structures; in the case discussed here, the trench between the nanotube 
contacts was 500\,nm wide and 220\,nm deep. For characterization and selection, 
the devices are only tested electrically at room temperature. The low 
temperature transport experiments with suitable samples are conducted at 
$T_{\text{base}}=300\un{mK}$ in a helium-3 evaporation cooling system. Here, in 
addition to a typical Coulomb blockade measurement setup \cite{kouwenhoven}, a 
radio frequency signal for mechanical excitation is applied contact-free by an 
antenna nearby in the cryostat \cite{highq}. Figure \ref{probe}(b) depicts a 
schematic of this measurement setup: we apply a bias voltage and measure the 
resulting current; a gate voltage is used for varying the electrochemical 
potential and thereby also the charge of the carbon nanotube.

\section{Basic electronic characterization}
\begin{figure*}[htb]%
\includegraphics*[width=\textwidth]{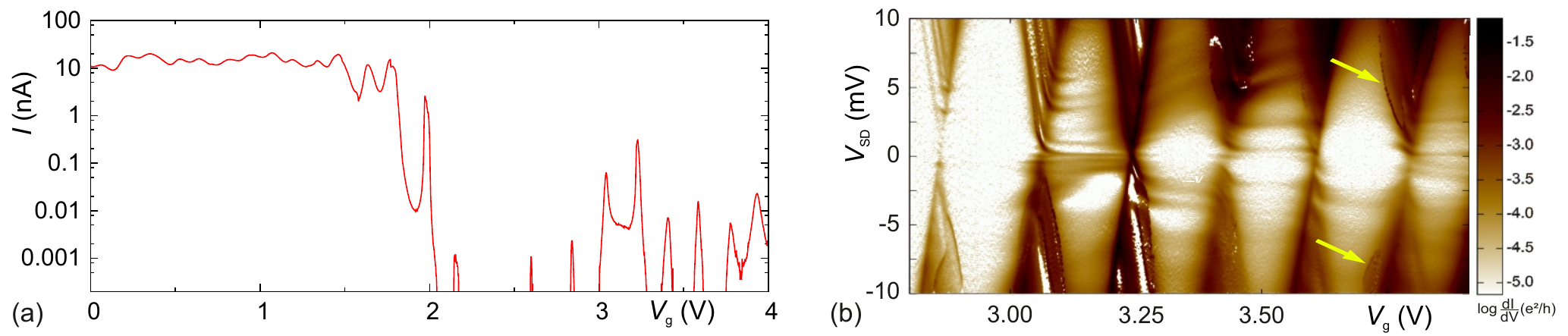}
\caption{
(a) dc current as function of applied gate voltage \vg\ for constant 
$\vsd=2\un{mV}$, at $T\simeq 300\un{mK}$. The device displays transparent 
behaviour in the hole regime ($\vg \lesssim 2.2\un{V}$), a small bandgap, and 
irregular Coulomb oscillations in electron conduction ($\vg \gtrsim 
2.6\un{V}$). (b) Differential conductance \didv\ (numerically obtained from a 
dc-current measurement) as function of gate voltage \vg\ and bias voltage \vsd, 
in the few-electron regime (logarithmical color scale). The yellow arrows mark 
regions of nano-electromechanical feedback 
\cite{strongcoupling,magdamping,Usmani2007Strong}.
}
\label{elektronik}
\end{figure*}
Figure \ref{elektronik}(a) displays a low-temperature ($T\simeq 300\un{mK}$) 
measurement of the current $I(\vg)$ through our carbon nanotube sample as a 
function of applied gate voltage \vg, for constant $\vsd=2\un{mV}$. 
The sample displays the typical behaviour of a small bandgap carbon nanotube. 
On the hole conduction side ($\vg \lesssim 2.2\un{V}$) we only observe few 
oscillations of the current and a subsequent fast transition into an open 
transport regime \cite{Liang}. Note, however, that the overall resistance 
remains high ($R\simeq 180\un{k\text{$\Omega$}}$ at $\vg=0$), indicating either 
a high series resistance in the leads or a multi-dot structure. On the electron 
conduction side ($\vg \gtrsim 2.6\un{V}$) we observe sharp Coulomb oscillations, 
however no clear shell structure can be observed. By additionally varying the 
bias voltage \vsd\ we obtain the typical ``Coulomb diamond'' stability diagram, 
see Fig. \ref{elektronik}(b). Multiple, strongly gate-dependent inelastic 
cotunnelling lines without clear pattern can be observed 
\cite{PhysRevLett.86.878,Goss}, hinting at a potential structure more complex 
than a single minimum. Also in nonlinear transport no regular shell structure 
can be observed. Of note in the measurement of Fig.~\ref{elektronik}(b), 
however, are the rounded regions marked with arrows -- here, electromechanical 
feedback leads to self-driving of the mechanical motion in absence of an 
external rf driving signal \cite{strongcoupling,magdamping,Usmani2007Strong}.

\section{Driven mechanical resonator measurements}
\begin{figure*}[htb]%
\includegraphics*[width=\textwidth]{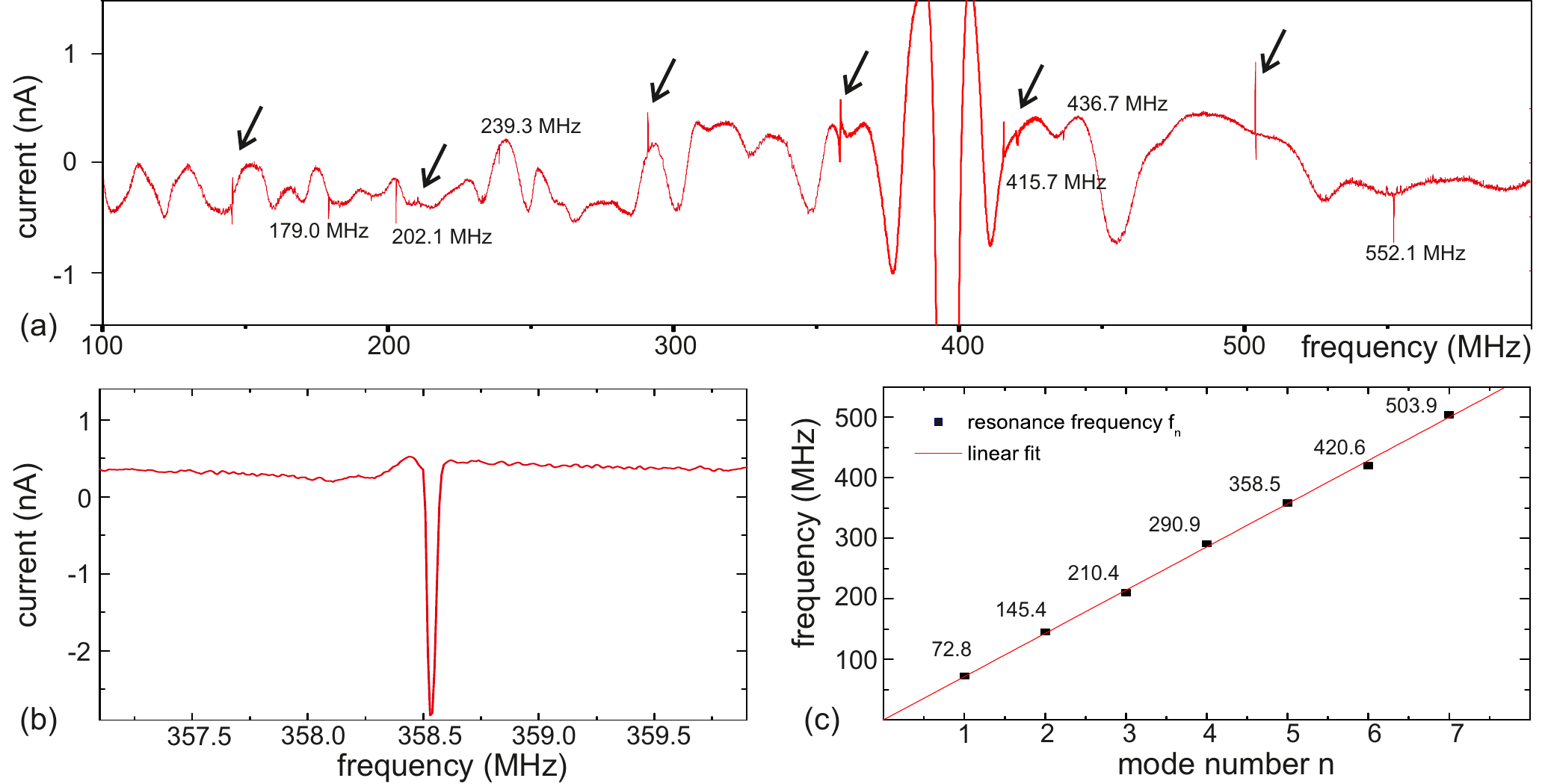}
\caption{(a) Time-averaged dc current through the carbon nanotube as function 
of the frequency of the rf driving signal, for constant $\vsd=2\un{mV}$, 
$\vg=2.234\un{V}$, and nominal rf generator power $P=2\un{dBm}$. Black arrows 
mark the resonance features used in the plot of (c); additionally observed 
resonances are marked with their frequency. (b) Detail from (a) (higher 
resolution measurement): exemplary resonance trace around $f=358.5\un{MHz}$. 
The peak width corresponds to a quality factor $Q=11800$. (c) Selected 
resonance frequencies from measurements as in (a) as function of assigned mode 
number. A sequence of harmonics can be observed; the linear fit results in 
$f_{\text{fit},n}=n\; (71.4 \pm 0.4) \un{MHz}$.}
\label{mechanik}
\end{figure*}
A mechanical resonance detection measurement is shown in Figure 
\ref{mechanik}(a). The bias voltage $\vsd=2\un{mV}$ and the gate voltage 
$\vg=3.234\un{V}$ are kept constant, the frequency of an applied rf-signal 
(compare Fig.~\ref{probe}(b)) is varied across a large range at constant 
nominal generator power $P=2\un{dBm}$. Since in this setup the cabling used for 
the radio-frequency signal is not particularly optimized, the actual power 
transmitted to the sample varies over the observed frequency range, leading to 
the large-scale oscillatory behaviour in Fig.~\ref{mechanik}(a). Mechanical 
resonances of the carbon nanotube show up as sharp spikes in the recorded 
current (for details of the detection mechanism, see e.g.\ \cite{highq}). We 
observe several resonance frequencies ranging from 72.8\,MHz to 552.1\,MHz. 
Fig.~\ref{mechanik}(b) displays an exemplary detail zoom measurement of the 
resonance in Fig.~\ref{mechanik}(a) around $f=358.5\un{MHz}$. The width of the 
measured peak corresponds to a quality factor of $Q=f/\Delta f= 11800$; maximum 
quality factors observed on this device were $Q_{\text{max}}\simeq 72000$ at 
$T\simeq 300\un{mK}$.

Figure \ref{mechanik}(c) shows selected extracted peak positions plotted as a 
function of an assigned mode number of the mechanical carbon nanotube resonator. 
The peaks form a sequence of approximately integer multiple frequencies; the 
dependence of the resonance frequency on the mode number can be fitted 
linearly, leading to a good agreement for $f_{\text{fit},n}=n\; (71.4 \pm 0.4) 
\un{MHz}$. This indicates the presence of a string under tension, since for the 
case of a hanging chain, where the bending rigidity dominates mechanical 
behaviour, higher vibration modes do not occur at integer multiples of the 
fundamental frequency \cite{book-cleland,nl-witkamp:2904}. The observation is 
consistent with recent measurements on short nanotube segments 
\cite{doi:10.1021/nl203279v}. 

In addition, in the plot of Fig.~\ref{mechanik}(a) further resonances appear. 
These are each marked in the plot with the resonance frequency. While a 
mechanical origin of these features is likely because of the sharp frequency 
response, a detailed identification cannot be made from our data.

\begin{figure}[htb]%
\includegraphics*[width=\columnwidth]{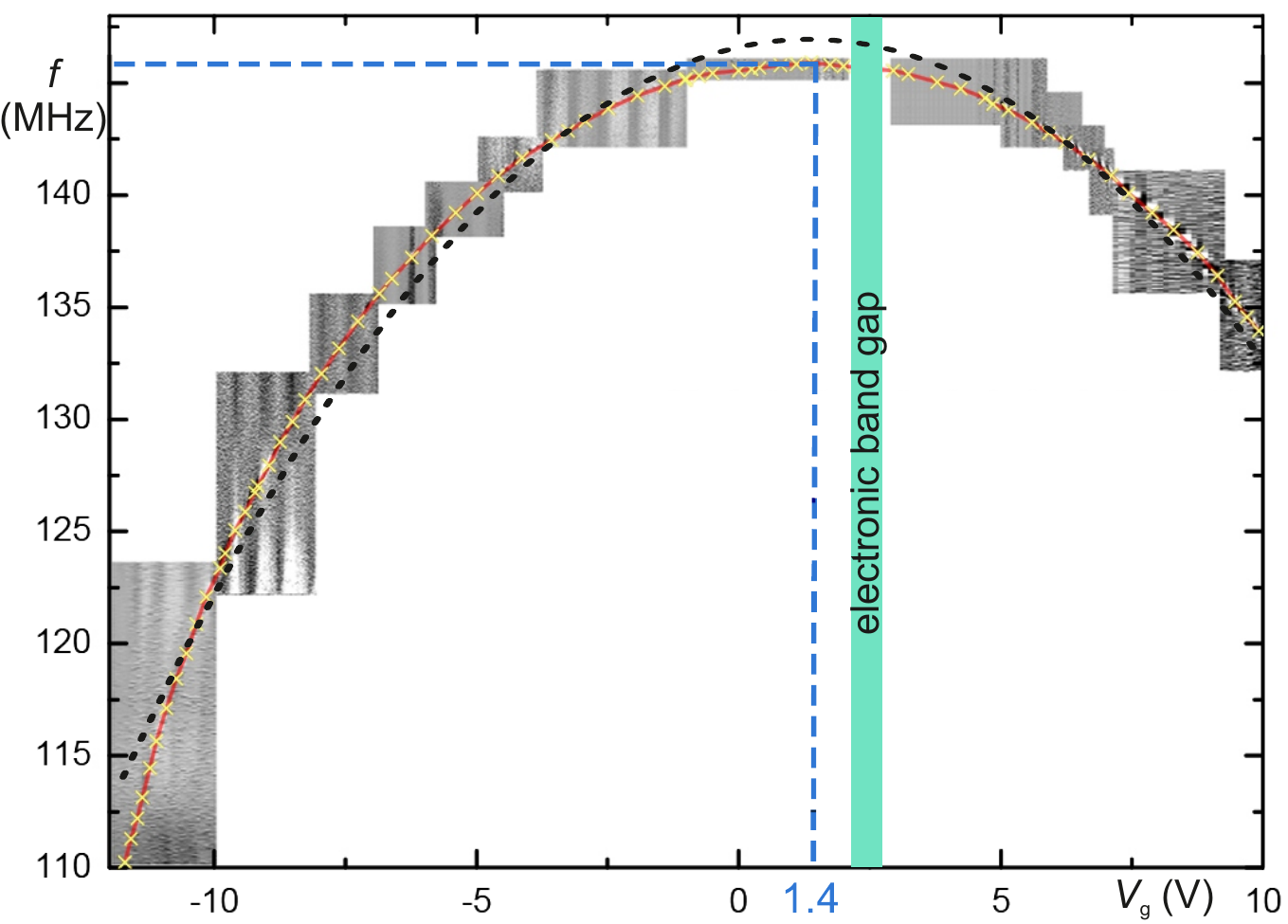}
\caption{Background: numerical derivative $\textrm{d}I/\textrm{d}f$ of the 
measured time-averaged dc current as function of external rf driving frequency 
$f$ and gate voltage \vg, for nominal driving power $P=0\un{dBm}$ and bias 
voltage $\vsd=2\un{mV}$. The extracted resonance peak positions for the first
harmonic frequency $f_2$ as function of gate voltage \vg\ are overlaid as 
yellow crosses. The black dotted line corresponds to a parabolic fit, see 
equation \ref{parabola}.} 
\label{resonanzkurve}
\end{figure}
\section{Gate voltage dependence of mechanical resonance}
Varying both the driving frequency $f$ and the gate voltage \vg\ enables to 
trace the resonant features in a 2D plot. This is done in Fig. 
\ref{resonanzkurve} for a wide gate voltage range from -12\,V to 10\,V. 
We have chosen the first harmonic frequency for this evaluation since it 
provides the best signal to noise ratio. As already stated, the expected 
behaviour would be an increase of the mechanical resonance frequency for 
increasing gate voltages. Fig.~\ref{resonanzkurve}, however, displays a strong 
negative curvature of the resonance frequency; at $\vg=-12\un{V}$ far in the 
hole conduction regime the resonance frequency is reduced to approximately 75\% 
of its maximum value. For electron conduction, the same effect emerges 
symmetrically.

A decrease of resonance frequency has been observed previously in literature in 
measurements on a suspended metallized SiC beam and on a doubly clamped InAs 
nanowire resonator \cite{apl-kozinsky:253101,prb-solanki:115459} and also in 
carbon nanotube mechanical resonators \cite{HighQuality}. There, it is 
explained via so-called electrostatic softening of the vibration. For this 
effect, an out-of-plane motion of the carbon nanotube towards and away from the 
backgate is required, since the capacitance between resonator and gate 
electrode has to vary. 

Assuming that the built-in tension in the carbon nanotube device is dominant 
at low gate voltage \vg, we approximate that the mechanical tension and thereby 
the purely mechanical spring constant does not change in the observed range of 
\vg. Following 
\cite{HighQuality}, we can then approximate 
\begin{equation}\label{parabola}
f(\vg)=f_{\text{max}} -\beta (\vg - V_{g,0})^2, 
\end{equation}
where we define with $m$ the mass and $L$ the length of the nanotube
\begin{equation}
f_{\text{max}}=\frac{1}{2}\sqrt{\frac{T_0}{m L}}. 
\end{equation}
The curvature of the parabola $f(\vg)$ is connected to the second derivative of 
the capacitance $C_g$ between gate and resonator as function of the distance 
$h$ between them, $C_g''(h)=\text{d}^2 C_g/\text{d}h^2$, via
\begin{equation}
\beta=f_{\text{max}}\frac{C_g'' L}{4\pi^2 T_0}.
\end{equation}

The black dotted line in Fig.~\ref{resonanzkurve} corresponds to a parabolic 
fit using Eq.~\ref{parabola} with the parameters $f_{\text{max}}=146.9\un{MHz}$, 
$V_{g,0}=1.4\un{V}$, and $\beta=0.192\un{MHz/V$^2$}$. Assuming a mass of the 
carbon nanotube $m=0.17\times 10^{-21}\un{kg}$ and a nanotube length equal the 
trench width $L=500\un{nm}$, these values lead to $T_0=7.3\un{pN}$ and 
$C_g''=7.5\times 10^{-7}\un{F/m$^2$}$. 

As can be seen in Fig.~\ref{resonanzkurve}, the parabolic fit does not 
accurately represent the functional dependence of $f(\vg)$. Several possible 
mechanisms can contribute here. Eq.~\ref{parabola} assumes that the charge on 
the nanotube is proportional to the applied gate voltage, which in particular 
does not hold within and close to the electronic band gap. In addition the 
mechanical tension varies, leading to additional contributions to $f$.

As a consistency check, we calculate the distance between carbon nanotube 
and gate required to obtain this value of $C_g''$, using the simple model of a 
thin beam with radius $r$ above an infinite conductive plane. Assuming a 
nanotube radius $r=2\un{nm}$, and $L$ and $C_g''$ from above, we obtain a 
required distance in vacuum between nanotube and gate of $h=1\un{$\mu$m}$. In 
spite of the many approxmations used, this result is indeed of the correct 
order of magnitude, compared to a trench depth below the carbon nanotube of 
$220\un{nm}$ and an additionally remaining silicon oxide layer of $380\un{nm}$.

\section{Conclusion}
We measure electronic and mechanical properties of a high-quality factor 
carbon nanotube vibrational resonator at cryogenic temperatures. Both quantum 
dot behaviour and multiple driven mechanical resonances are observed. A 
sequence of higher harmonics occurs approximately at multiple integers of a 
base frequency, indicating that the nanotube resonator is under 
tension. The gate voltage dependence of the resonance frequency of the first 
harmonic $f_2(\vg)$ displays a distinct negative curvature; the frequency can 
be tuned from a maximum $f_2(V_g=1.4\un{V})=145.9\un{MHz}$ down to 
$f_2(V_g=-11.7\un{V})=110.2\un{MHz}$. We successfully model this by 
electrostatic softening.

The authors would like to thank the Deutsche Forschungsgemeinschaft (Emmy 
Noether grant Hu 1808/1-1, SFB 631 TP A11, GRK 1570) and the Studienstiftung 
des deutschen Volkes for financial support.

\bibliography{paper}

\end{document}